\documentclass{emulateapj}
\usepackage{amssymb,amsmath,amsthm}
\usepackage{longtable}

\def\Msun{$M_{\odot}$}
\def\Zsun{$Z_{\odot}$}

\newcommand{\MgasSZ}{\mbox{$M_{\mbox{\scriptsize gas,SZ}}$}}
\newcommand{\MgasX}{\mbox{$M_{\mbox{\scriptsize gas,X-ray}}$}}

\newcommand{\xmm}{\it XMM-Newton\rm}
\newcommand{\chandra}{\it Chandra\rm}

\newcommand{\rms}{{\it rms}}

\def\ho{$H_0$}
\def\kmsmpc{km~s$^{-1}$~Mpc$^{-1}$}

\def\xmmj{XMMXCS~J2215.9-1738}
\def\xmmjb{XMMU~J2235-2557}
\def\xmmjc{2XMM J083026.2+524133}
\def\clA{Cl\,J1415.1+3612}

\def\jkcs{JKCS~041}
\def\iscs{ISCS1438.1+3338}
\def\rxja{RX~J0848+4453}
\def\rxjb{RX~J0849+4452}
\def\rxjd{RX~J0910+5422}
\def\rxje{RX~J1252-2927}
\def\sparcsa{SpARCSJ1638}

\def\sza{SZA\rm}

\shorttitle{SZA observations of $z\geq1$ galaxy clusters}
\shortauthors{Culverhouse et al.}

\tabletypesize{\scriptsize}
\begin{document}

\title{Galaxy Clusters at $z\geq 1$: Gas Constraints from the 
Sunyaev-Zel'dovich Array}
 %z=1.39, and observations of other $z\geq 1$ galaxy clusters with
 %the Sunyaev-Zel'dovich Array}

\author{
T.~L.~Culverhouse,\altaffilmark{1,2}
M.~Bonamente,\altaffilmark{3,4}
E.~Bulbul,\altaffilmark{3}
J.~E.~Carlstrom,\altaffilmark{1,2,5,6}
M.~B.~Gralla,\altaffilmark{1,2}
C.~Greer,$\!$\altaffilmark{1,2}
N.~Hasler,\altaffilmark{3} 
D.~Hawkins,\altaffilmark{7}
R.~Hennessy,\altaffilmark{1,2} 
N.~N.~Jetha,\altaffilmark{3,4}
M.~Joy,\altaffilmark{4}
J.~W.~Lamb,\altaffilmark{7}
E.~M.~Leitch,\altaffilmark{1,2} 
D.~P.~Marrone,\altaffilmark{1,5,8}
A.~Miller,\altaffilmark{9,10,11}
T.~Mroczkowski,\altaffilmark{9,12,13}
S.~Muchovej,\altaffilmark{7,13}
C.~Pryke,\altaffilmark{1,2,5}
M.~Sharp,\altaffilmark{1,6} 
D.~Woody,\altaffilmark{7}
S.~Andreon,\altaffilmark{14}
B.~Maughan,\altaffilmark{15}
and S.~A.~Stanford, \altaffilmark{16,17}}

\altaffiltext{1}{Kavli Institute for Cosmological Physics, University of Chicago, Chicago, IL 60637}
\altaffiltext{2}{Department of Astronomy and Astrophysics, University of Chicago, Chicago, IL 60637}
\altaffiltext{3}{Department of Physics, University of Alabama, Huntsville, AL 35899}
\altaffiltext{4}{Space Sciences - VP62, NASA Marshall Space Flight Center, Huntsville, AL 35812}
\altaffiltext{5}{Enrico Fermi Institute, University of Chicago, Chicago, IL 60637}
\altaffiltext{6}{Department of Physics, University of Chicago, Chicago, IL 60637}
\altaffiltext{7}{California Institute of Technology, Owens Valley Radio Observatory, Big Pine, CA 93513} 
\altaffiltext{8}{Hubble Fellow}
\altaffiltext{9}{Columbia Astrophysics Laboratory, Columbia University, New York, NY 10027}
\altaffiltext{10}{Department of Physics, Columbia University, New York, NY 10027}
\altaffiltext{11}{Alfred P. Sloan Fellow}
\altaffiltext{12}{Department of Physics and Astronomy, University of Pennsylvania, Philadelphia, PA 19104}
\altaffiltext{13}{Department of Astronomy, Columbia University, New York, NY 10027}
\altaffiltext{14}{INAF-Osservatorio Astronomica di Brera, via Brera 28, 20121, Milano, Italy}
\altaffiltext{15}{HH Wills Physics Laboratory, University of Bristol, Tyndall Ave, Bristol BS8 1TL, UK.}
\altaffiltext{16}{University of California, Davis, CA 95618}
\altaffiltext{17}{Institute of Geophysics and Planetary Physics, Lawrence Livermore National Laboratory, Livermore, CA 94550}

\begin{abstract}

We present gas constraints from Sunyaev-Zel'dovich (SZ) effect measurements in a sample of eleven 
X-ray and infrared (IR) selected galaxy clusters at $z\geq$1, using data from
 the Sunyaev-Zel'dovich Array (\sza).
The cylindrically integrated Compton-$y$ parameter, $Y$, is calculated by fitting the data
to a two-parameter gas pressure profile. 
Where possible, we also determine the temperature of the hot intra-cluster plasma from 
\chandra\ and \xmm\ data, and constrain the gas mass within the same aperture ($r_{2500}$)
as $Y$. 
The SZ effect is detected in the clusters for which the X-ray data indicate gas masses above 
$\sim10^{13}$\Msun, including \xmmjb\ at redshift $z=1.39$, which to date is one of the most 
distant clusters detected using the SZ effect.
None of the IR-selected targets are detected by the \sza\ measurements, indicating low gas
masses for these objects.
For these and the four other undetected clusters, we quote upper limits on $Y$ and $M_{gas,SZ}$, 
with the latter derived from scaling relations calibrated with lower redshift clusters.
We compare the constraints on $Y$ and X-ray derived gas mass \MgasX\ to self-similar 
scaling relations between these observables determined from observations of lower redshift 
clusters, finding consistency given the measurement error.

\end{abstract}


\section{Introduction}

The Sunyaev-Zel'dovich (SZ) effect is a distortion in the spectrum of the cosmic microwave background
 (CMB) radiation caused by inverse Compton scattering of CMB 
photons with the ionized gas in galaxy clusters \citep{sunyaev1972,birkinshaw1991,carlstrom2002}. 
The amplitude of the SZ effect is commonly described by the Compton $y$-parameter, which
for a thermal population of electrons is given by the 
integral of the gas pressure along the line of sight through the cluster:
\begin{equation}
\label{eq:y}
y=\int \frac{\sigma_{\rm T}}{m_ec^2}P_{e} dl
\end{equation}
\noindent In this expression, $\sigma_{\rm T}$ is the Thomson scattering cross 
section, $m_{e}$ is the electron mass, $c$ is the speed of light, 
 $P_{e}$ is the gas pressure, and the integration is along the line-of-sight.
Integrating $y$ over the solid angle $\Omega$ yields the integrated Compton 
parameter $Y$, which is proportional to the thermal energy 
of the cluster \citep[][hereafter B08]{motl2005, bonamente2008}.

This paper reports observations of $z\geq 1$ clusters made with the Sunyaev-Zel'dovich
Array (\sza), and aims to provide constraints on the gas properties of the clusters, and
a comparison to existing scaling relations.
~In Section~\ref{sec:sample} we describe the sample of clusters, 
Section~\ref{sec:szadata} describes the collection and analysis of
the \sza\ data, and Section~\ref{sec:xrays} presents an analysis of cluster 
X-ray data (where available) from the \chandra\ and \xmm\ observatories.
The results and discussion, including a comparison of SZ and X-ray cluster gas 
properties, are given in Section~\ref{sec:results}.
Throughout this document we use the cosmological parameters \ho=73~\kmsmpc, $\Omega_m=0.27$
and $\Omega_{\Lambda}=0.73$.
Unless otherwise stated, all uncertainties correspond to the 16\% and 84\% 
percentiles of the probability distribution function (68\% confidence interval).

\begin{deluxetable*}{lcccl|cccc|cc}[!ht]
\tablecaption{Cluster Sample \label{tab:ubertable}}
\tablehead{
\colhead{Cluster} & 
\colhead{$z$}     &
\colhead{$R.A.$}  &  
\colhead{$decl.$} &
\colhead{Detection, (Ref.)\tablenotemark{i}} &
\colhead{$\rm{t_{int}}$\tablenotemark{a}} &
\multicolumn{3}{c}{\underline{Short Baselines\tablenotemark{b} } \hspace{2cm}} &
\colhead{$D_{A}^{2}Y$\tablenotemark{g}} & 
\colhead{$M_{gas,SZ}$\tablenotemark{h}} \\
\colhead{} & 
\colhead{} & 
\colhead{} & 
\colhead{} & 
\colhead{} & 
\colhead{} & 
\colhead{FWHM\tablenotemark{c}} & 
\colhead{$\sigma$\tablenotemark{d}} &
\colhead{B\tablenotemark{e}} &
\colhead{} & 
\colhead{} \\
\colhead{} & 
\colhead{} & 
\colhead{} & 
\colhead{} & 
\colhead{} & 
\colhead{(hrs)} &
\colhead{(arcsec)} &
\colhead{(mJy)} &
\colhead{($\mu$K)} &
\colhead{($10^{-5}Mpc^{2}$)}   & 
\colhead{($10^{13}M_{\odot}$)} 
}
\startdata
\jkcs    & 1.90 & 02 26 44 & -04 41 37 & IR, (1)     & 30.6                  & 86.6 $\times$96.7  & 0.13  & 19.9 & $<0.68$                & $<0.42$                 \\  
\xmmjc   & 0.99 & 08 30 26 & +52 41 33 & X-ray, (2)  & 23.3                  & 81.7 $\times$109.3 & 0.17  & 24.4 & $2.01^{+0.34}_{-0.32}$ & $1.12^{+0.25}_{-0.25}$  \\ 
\rxja    & 1.27 & 08 48 35 & +44 53 49 & IR, (3)     & 44.5                  & 81.4 $\times$111.6 & 0.11  & 15.5 & $<0.18$                & $<0.28$                 \\ % MASS NOT RIGHT
\rxjb    & 1.26 & 08 49 58 & +44 51 55 & X-ray, (4)  & 25.5                  & 82.3 $\times$110.7 & 0.15  & 21.1 & $<0.77$                & $<0.53$                 \\ 
\rxjd    & 1.11 & 09 10 44 & +54 22 09 & X-ray, (5)  & 19.1                  & 83.6 $\times$107.1 & 0.17  & 24.3 & $<0.29$                & $<0.39$                 \\ % MASS NOT RIGHT
\rxje    & 1.24 & 12 52 54 & -29 27 17 & X-ray, (6)  & 12.2                  & 97.8 $\times$166.9 & 0.28  & 22.0 & $<1.18$                & $<0.72$                 \\ 
\clA     & 1.03 & 14 15 11 & +36 12 03 & X-ray, (7,8)& 55.3\tablenotemark{f} & 95.9 $\times$118.1 & 0.12  & 13.6 & $2.39^{+0.57}_{-0.56}$ & $1.26^{+0.36}_{-0.37}$  \\ 
\iscs    & 1.41 & 14 38 09 & +34 14 19 & IR, (9)     & 17.5                  & 110.9$\times$129.1 & 0.21  & 18.8 & $<0.36$                & $<0.42$                 \\ % MASS NOT RIGHT
\sparcsa & 1.20 & 16 38 52 & +40 38 43 & IR, (10)    & 36.0                  & 78.9 $\times$108.1 & 0.13  & 19.5 & $<0.70$                & $<0.50$                 \\ 
\xmmjb   & 1.39 & 22 35 21 & -25 57 42 & X-ray, (11) & 42.1                  & 103.7$\times$150.8 & 0.14  & 11.5 & $1.87^{+0.34}_{-0.33}$ & $0.96^{+0.24}_{-0.24}$  \\ 
\xmmj    & 1.46 & 22 15 58 & -17 38 03 & X-ray, (12) & 10.8                  & 107.8$\times$130.1 & 0.25  & 22.8 & $<0.32$                & $<0.38$                 \\ % MASS NOT RIGHT 
\enddata

\tablenotetext{a}{On source integration time, unflagged data.}
\tablenotetext{b}{Short baselines correspond to (0-2k$\lambda$).}
\tablenotetext{c}{Synthesized beam approx. FWHM.}
\tablenotetext{d}{Achieved \rms \ noise in short baseline maps.}
\tablenotetext{e}{Corresponding brightness sensitivity in short baseline maps.}
\tablenotetext{f}{34.1 hours of `V-array' integration from \citet{muchovej2007}; 21.2 hours from `L-array'.}
\tablenotetext{g}{$Y$ constraints from \sza\ data. Where there are sufficient X-ray counts to determine $r_{2500}$ (see Table~\ref{tab:xray}), 
                  $Y$ is calculated within this radius; otherwise, an angular aperture of radius 30 arcsec is used. Upper limits are calculated at 95\% confidence.}
\tablenotetext{h}{\MgasSZ\ determined from $Y$ constraints and scaling relations from~\cite{bonamente2008}, and are 
                  independent of the X-ray determined gas mass \MgasX.}
\tablenotetext{i}{Cluster references: (1) \cite{andreon2008}; (2) \cite{lamer2008}; (3) \cite{stanford1997}; 
               (4) \cite{rosati1999}; (5) \cite{stanford2002}; (6) \cite{rosati2004}; (7) Redshift from \cite{maughan2006}; 
               (8) \cite{perlman2002}; (9) \cite{stanford2005}; (10) \cite{muzzin2009}; (11) \cite{mullis2005}; (12) \cite{stanford2006}.}

\end{deluxetable*}

\section{Sample selection}
\label{sec:sample}
We obtained \sza\ observations of an ad hoc sample of eleven clusters with 
$z\geq$1 discovered in either X-ray or infrared (IR) imaging surveys --- basic information
 about the clusters are given in Table~\ref{tab:ubertable}.
Surveys in these bands can yield large numbers of high redshift cluster candidates 
using a variety of methods.
These include the red-sequence in the optical~\citep{gladders2000,gladders2005}, 
its extension into the IR~\citep{stanford2005,andreon2008,eisenhardt2008,muzzin2009}, and imaging 
surveys and/or serendipitous detection in X-rays~\citep[e.g.][]{stanford2006,lamer2008}.

X-ray observations provide direct evidence for the hot plasma which 
typically constitutes $\sim10\%$ of the cluster total mass;
seven clusters in the sample are the most massive X-ray detected systems at
redshift above one.
Since this is the same plasma which causes the SZ effect, clusters in the sample 
originally detected in X-rays are expected to have significant 
SZ signal, provided they are of sufficient gas mass and temperature.

The IR-detected clusters in the sample were selected as optically rich 
candidates with properties typical of massive clusters:
\iscs\ was detected in the Spitzer/IRAC Shallow Survey as an overdensity
 of galaxies with photometric redshifts between $1.3<z<1.5$; member 
galaxies were confirmed with Keck optical spectroscopy to have 
$\Delta z=0.01$. 
The most massive spectroscopically confirmed cluster from the SpARCS North 
Survey is \sparcsa, with initial detection via two-filter imaging.
\jkcs\ was discovered using a modified red-sequence method applied
 to J and K band data in the UKIRT Infrared Deep Sky Survey; 
\chandra\ follow-up revealed a low luminosity, diffuse source
 of X-ray emission at the cluster location, with a photometric
redshift of 1.9.
Also discovered in the IR, member galaxies of \rxja\ exhibit very
red J-K colors, with follow up spectrocopy confirming the member
redshifts are within $\Delta z=0.002$ of each other. 
This cluster has also been observed in the X-ray with \chandra;
~\cite{santos2008} present a recent analysis.

The most massive, high redshift cluster candidates in these different 
surveys provide a starting point for studies of the SZ effect in 
galaxy clusters at $z\geq 1$.
Constraints on $Y$ from the SZ effect alone provide useful information on the
presence of hot gas, while joint analysis with X-ray data allows 
comparison with more local samples via scaling relations, to test for 
evolutionary effects.

\section{Sunyaev-Zel'dovich Effect Analysis}
\label{sec:szadata}

\subsection{SZA Observations}
\label{subsec:obs}

\begin{figure*}[!ht]
\resizebox{\textwidth}{!}{
\includegraphics[width=1.5in]{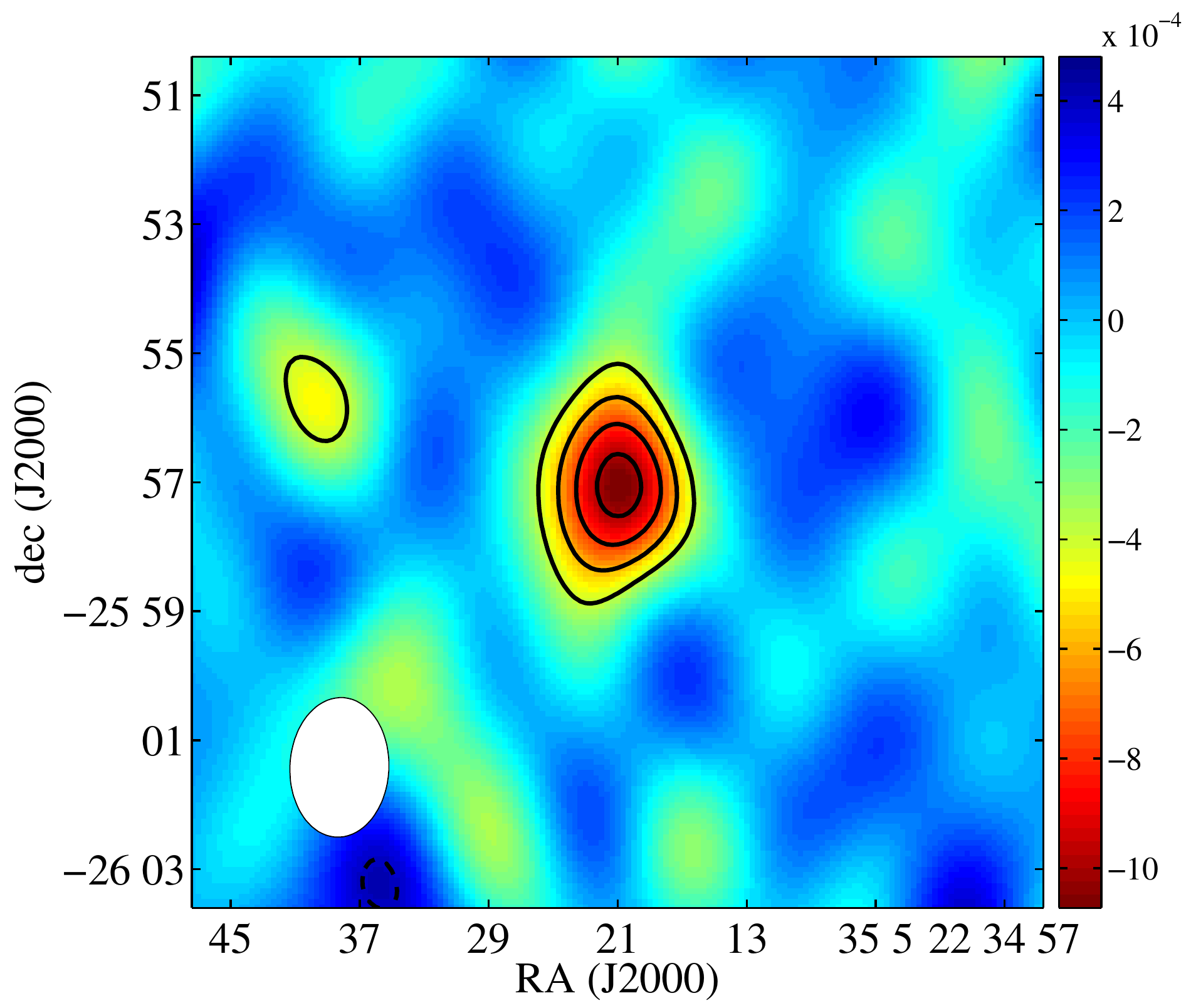}   \\
\includegraphics[width=1.5in]{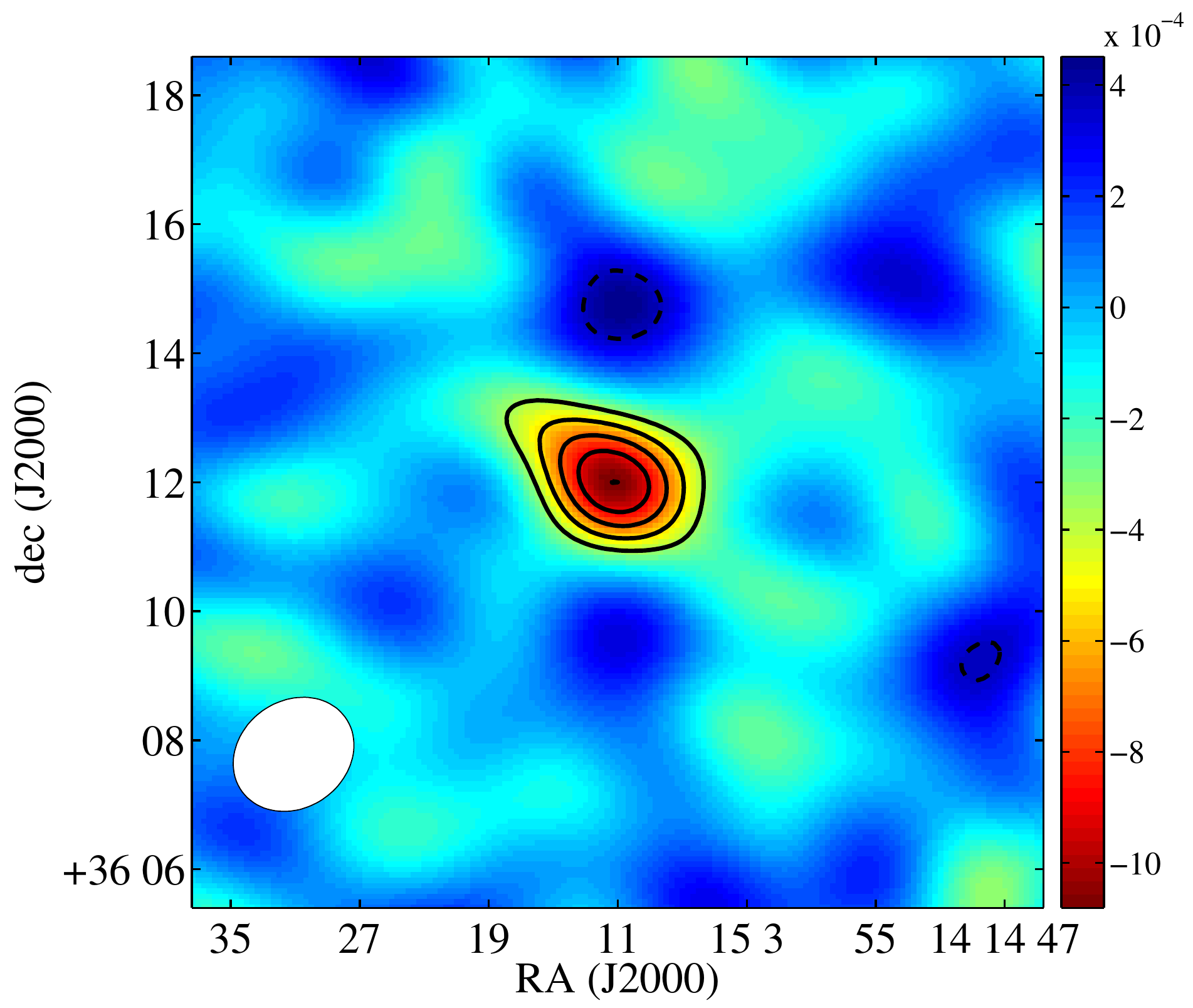}     \\ 
\includegraphics[width=1.5in]{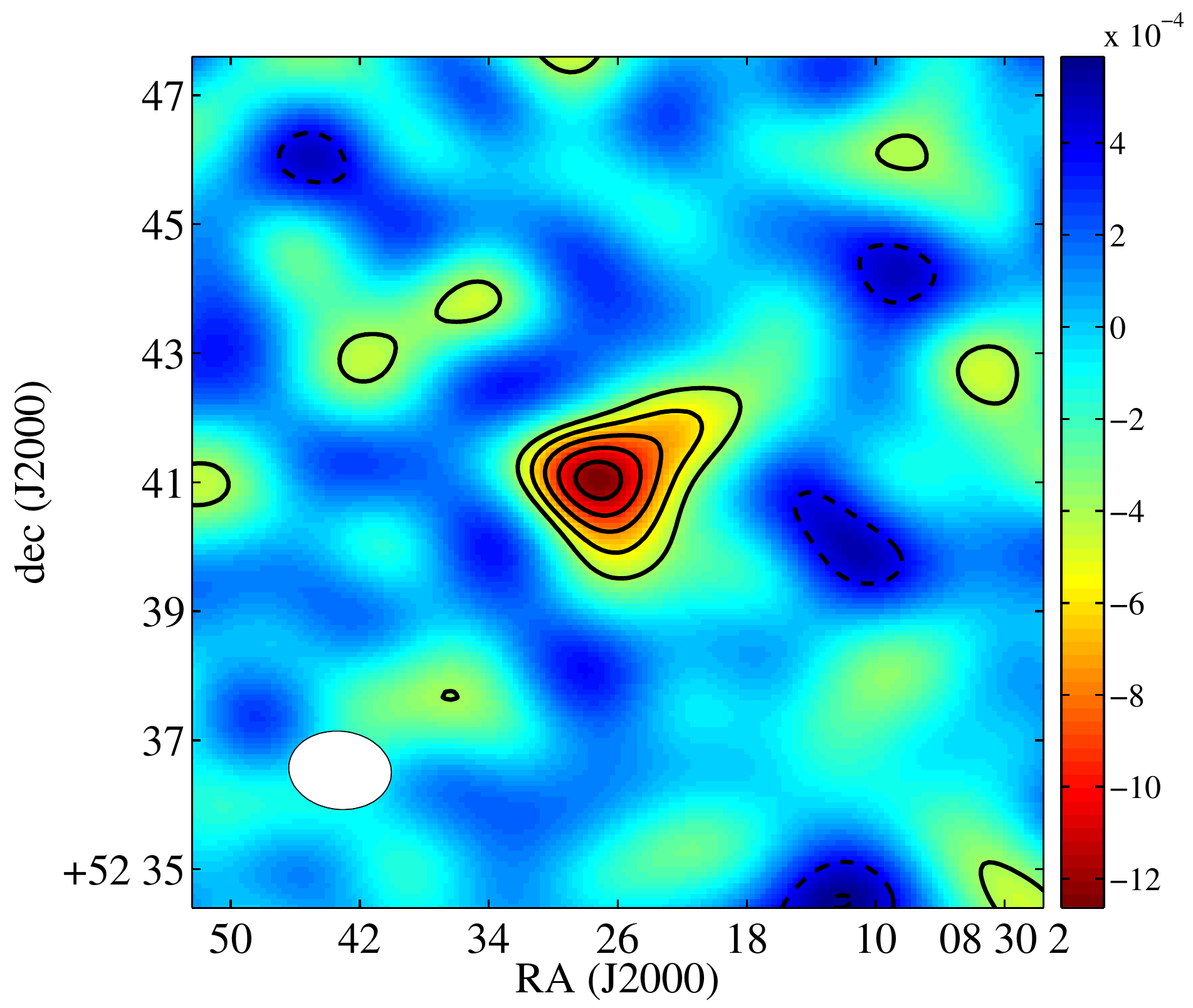}   \\
}
\caption{CLEANed \sza\ short ($<$2k$\lambda$) baseline images of the three
clusters with significant SZ effect detections: Left to right, \xmmjb\, \clA\ and
\xmmjc, with the colorscale in Jy/beam. The contours begin at 
$2\sigma$ and are spaced at unit intervals of the map rms $\sigma$. 
The white ellipse represents the half-power point of the elliptical gaussian
that approximates the main lobe of the synthesized beam.
Radio sources have been %CLEANed 
removed
for display purposes.}
\label{fig:xmmj}
\end{figure*}

The \sza\ is an interferometric array comprising eight 3.5-meter telescopes, 
capable of observations in an 8-GHz-wide band centered on
31~GHz or 90~GHz.
The instrument was configured to operate at 31~GHz for the observations 
reported here --- see \citet{muchovej2007} for further details.
The field of view of the \sza\ is given by the primary beam of a single telescope,
 of FWHM approximately 11\arcmin\ at the center of the 31~GHz band, with
typical system temperatures $\sim$40K at this frequency.

Observations of each cluster were performed with different array configurations.
All observations of \iscs\ and two thirds of the tracks on \clA\ were observed from the 
OVRO valley floor site, using a compact six-telescope plus two outrigger array, 
 denoted `V' array here. 
\jkcs\ was initially observed for twelve days at the CARMA site.
An imaging array configuration (denoted `I') was used, with no outriggers but higher 
sensitivity to typical cluster angular scales.
A further eight days of data were taken in a `low dec' plus outrigger array, 
or `L' array. 
This configuration is similar to `V' array, but with the array stretched North-South
 to prevent excessive shadowing for low declination clusters.
Tracks on all other clusters were taken in `L' array. 

Pairs of close-packed telescopes form short baselines (typically of order 4-20m or 0.4-2$k\lambda$), 
which are sensitive to the signal from clusters on angular scales of order $1'$.
The outrigger telescopes form long baselines (50m or 2-10$k\lambda$) between themselves and the 
close-packed antennas; these baselines allow measurement of contaminating
radio sources which could otherwise mask the SZ effect. 
The long baseline dirty map noise is typically $\sim0.2$mJy, with resolution $\sim20''$. 
Further details of the \sza\ observations presented here, including on-source
integration time and sensitivity, and effective resolution (the {\it synthesized beam\rm}) 
of the short baseline maps, are given in Table~\ref{tab:ubertable}.

\sza\ data are processed in a pipeline developed within the \sza\
collaboration for the reduction and calibration of interferometric data, 
described in detail in \citet{muchovej2007}.
The pipeline produces calibrated visibilities --- samples of the Fourier
 transform of the sky brightness distribution multiplied by the primary beam:
\begin{equation}
  \label{eq:vis}
V(u,v)  = \int_{-\infty}^{+\infty}\int_{-\infty}^{+\infty}A_N(l,m)I(l,m) e^{ -2\pi i [ul + vm]}{dl\,dm},
\end{equation}
\normalsize where $A_N(l,m)$ is the normalized antenna beam pattern, $I(l,m)$ is the 
sky intensity distribution, $u$ and $v$ are the baseline lengths 
projected onto the sky, %(as seen in Figure~\ref{fig:uvcov}, 
and $l$ and $m$ are direction cosines measured with respect to the $(u,v)$ axes.
Fourier transforming the visibility data gives the sky convolved with the 
synthesized beam, or the `dirty map'.

Figure~\ref{fig:xmmj} presents CLEANed images made from only the \sza\ short-baseline 
data of the clusters in the sample for which we detect the SZ effect at $>5\sigma$, i.e., \xmmjc, 
\clA, and \xmmjb. 
Radio sources have been identified and removed using the long and short 
baseline data.  
Unresolved radio sources were found within $1'$ of the cluster
 center in \clA, and \xmmjb.  
Resolved emission from two low redshift galaxies was detected: from NGC 5529 4.9$'$ from 
the \clA\ position, and from NGC 7314  7.8$'$ from the \xmmjb\ position. 
No cluster in the sample has more than three detected radio sources within the \sza\
field of view.
 
 %center in only two clusters in the sample.
 %maps of these clusters are shown in Figure~\ref{fig:xmmj}.

\begin{deluxetable*}{lllcccc}[!ht]
\tablecaption{\chandra\ and \xmm\ observations  \label{tab:chandra-data}}
\tablehead{
\colhead{Cluster}&
\colhead{Obs ID}&
\colhead{Detector}&
\colhead{Exposure time} & % FILTERED exposure time
\colhead{\# Source photons}&
\colhead{ nH Column Density}&
\colhead{}\\
\colhead{} & \colhead{} & \colhead{} & \colhead{(ks)} & \colhead{} & \colhead{($10^{20}$~cm$^{-2}$)}
}
\startdata
\xmmjc     & 0092800201                   & MOS1,PN      & 60.0, 55.9       & 2495\tablenotemark{b} & 4.2 \\ % this is XMM...
\clA       & 4163                         & ACIS-I       & 86.9             & 1395\tablenotemark{a}      & 1.1 \\
\rxjd      & 2227, 2452                   & ACIS-I       & 161.2            & 411\tablenotemark{a}       & 1.9 \\
\rxje      & 4198                         & ACIS-I       & 161.4            & 540\tablenotemark{a}       & 6.1 \\
\rxjb      & 1708, 927                    & ACIS-I       & 186.6            & 392\tablenotemark{a}       & 2.8 \\
\rxja      & 1708, 927                    & ACIS-I       & 186.6            & 144\tablenotemark{a}       & 2.8 \\
\xmmjb     & 6975, 6976, 7367, 7368, 7404 & ACIS-S       & 195.5            & 1532\tablenotemark{b}      & 1.5 \\
\xmmj      & 7919, 8566                   & ACIS-S       & 85.9             & 161\tablenotemark{b}       & 2.0 \\
           & 0106660601                   & MOS1,MOS2,PN & 81.0, 82.1, 60.0 & 
&  \\			         
	   & 0106660101			  & MOS1,MOS2,PN &  55.8, 53.1, 42.7&
&     \\
	   & 0106660201			  & MOS1,MOS2,PN & 35.7, 37.7, 24.5 & 686\tablenotemark{b}       & 2.0 \\
\jkcs      & 9368                         & ACIS-S       & 78.8             & 114\tablenotemark{b}       & 2.3 \\
\enddata											         
\tablenotetext{a}{0.7-7 keV band}
\tablenotetext{b}{0.5-7 keV band}
\tablenotetext{~}{Source photons are for a cluster-centric region of radius $<30$ arcsec, except \xmmjc\ and
\clA\ for which we use $<60$ arcsec.
In comparison with~\cite{andreon2008}, who find 223 source photons for \jkcs\ in a 60 arcsec aperture 
between 0.2-2 keV, we find 230 source photons in our 0.5-7 keV band in a 60 arcsec aperture.}
\end{deluxetable*}

\subsection{Constraints on the integrated SZ effect signal}
\label{sec:szaanal}
 
For each cluster, $Y$ is constrained by fitting a model to the data; the model $y$
 map is generated by integrating a gas pressure profile along the line of sight as 
in Equation~\ref{eq:y}. 
The $y$ map is multiplied by the primary beam, Fourier transformed, interpolated at 
the $(u,v)$ coordinates of the measured visibilities and the $\chi^{2}$ evaluated
 for the model against the data.
We use the spherically symmetric \citet{nagai2007} model which describes
the pressure as a function of radius $r$ as
\begin{equation}
P_e(r) = \frac{P_{e,i}}{(r/r_{p})^{c}
\left[1+(r/r_{p})^{a}\right]^{(b-c)/a}}.
\label{eq:n07}
\end{equation}
\noindent In Equation~\ref{eq:n07}, $P_{e,i}$ is the pressure normalization, $r_{p}$ is
the characteristic scale radius, and $a$, $b$ and $c$ are parameters
describing profile slopes at intermediate ($r\approx r_{p})$, outer ($r>r_{p}$)
 and inner ($r\ll r_{p}$) radii.
As in~\cite{mroczkowski2009}, the power-law indices of the pressure
model are held fixed at (a,b,c)=(0.9,5.0,0.4).

The Monte Carlo Markov chain (MCMC) method described in \citet{bonamente2004} 
is used to determine the probability distributions of the free cluster model 
parameters $ r_{p}$ and $P_{e,i}$; positions of clusters without a significant 
decrement are fixed to the values in Table~\ref{tab:ubertable}, but are 
otherwise variable.
Following~\cite{muchovej2007}, an elliptical gaussian is used to model the two 
resolved galaxies found in the sample (Section~\ref{subsec:obs}), 
while unresolved sources are described by one amplitude and two position 
parameters --- the typical rms on unresolved source position is of order 0.3 arcsec.
The free parameters of the cluster, resolved and unresolved source models are
 determined simultaneously with the MCMC method.
Accepted ($r_{p}$, $P_{e,i}$) parameter pairs in the MCMC analysis are  
used to calculate the cylindrically-integrated $Y$ parameter over the solid
angle of the cluster, via Equations~\ref{eq:y} and~\ref{eq:n07}.

For clusters with $\geq 300$ X-ray source photons, we measure $Y$ 
out to a radius of $r_{2500}$ as determined from the X-ray data 
(see Table~\ref{tab:xray}), where $r_{\Delta}$ is defined as the radius 
at which the mean cluster density falls to $\Delta$ times the critical density 
at the cluster redshift $\rho_{c}(z)$: 
\begin{equation}
\frac{4}{3} \pi \, \rho_c(z) \, \Delta \, r_{\Delta}^3 = M_{tot}(r_{\Delta}).
\label{rvir}
\end{equation}
The choice of $\Delta=2500$ allows our $Y$ parameters to be compared directly
to the scaling relations of B08, which were derived with
$Y(<r_{2500})$ for a large sample of low to intermediate redshift clusters.
A fixed angular aperture of radius 30\arcsec\ is used for clusters with $< 300$
X-ray photons, evaluating $Y(<30\arcsec)$ rather 
than $Y(<r_{2500})$ (such clusters are not compared to the scaling relations).

The mean ($D_{A}^{2}\bar{Y}$) and 68\% confidence intervals are calculated 
from the resultant probability distributions of $D_{A}^{2}Y$.
If $D_{A}^{2}\bar{Y}>3\sigma_{l}$, where $\sigma_{l}$ is the 14th percentile of
the distribution, we quote the mean and 68\% confidence interval for each cluster.
If $D_{A}^{2}\bar{Y}<3\sigma_{l}$, we quote the 95\% confidence upper limit on 
$D_{A}^{2}Y$.
The results are presented in Table~\ref{tab:ubertable}, along with the equivalent
 gas mass constraints calculated from the low redshift $Y-$\MgasX\ scaling 
relation of B08, which assumes self-similar evolution as $\mathrm{log_{10}}(YD_{A}^{2}E(z)^{-2/3})=A+B\mathrm{log_{10}}(M_{gas})$,
with $A=-23.25$ and $B=1.41$ taken from all clusters in their sample.
Errors on \MgasSZ\ include the uncertainty in the low redshift scaling relation 
parameters as well as the uncertainty in $Y$, but do not include errors introduced 
by geometric effects when performing the cylindrical integral or the intrinsic scatter
in $Y$ at fixed $M_{gas}$
\footnote[1]{The apparent difference between the SZ-derived gas mass for \clA\
presented here and that in~\cite{muchovej2007} is due to the different $r_{2500}$
used; their value is reproduced from an earlier analysis of \xmm\ data by~\cite{maughan2006}, 
compared to our \chandra\ derivation in Section~\ref{sec:xrays}. When the same $r_{2500}$
is used, the gas masses are consistent, as expected.}.

Note that these mass constraints are entirely independent of the \MgasX\ calculated
in the following X-ray analysis, and serve as the only gas mass estimates
available for clusters with insufficient X-ray data.

\section{X-ray observations and data analysis}
\label{sec:xrays}

\begin{deluxetable*}{lcc|ccc|cc}[ht]
\centering
\tablecaption{Image and Spectral analysis of the X-ray data \label{tab:xray}}
\tablehead{
\colhead{Cluster} &
\colhead{$kT$} &
\colhead{L$_{\rm X} $}&
\colhead{$n_{e0}$} &
\colhead{$r_{c}$} &
\colhead{$\beta$}&
\colhead{$r_{2500}$} &
\colhead{$\mathrm{M_{gas,X-ray}}$} \\%&
\colhead{} & \colhead{(keV)} & 
 \colhead{($10^{44} erg s^{-1}$)} & \colhead{$(10^{-2} cm^{-3})$} & \colhead{(arcsec)} & & \colhead{(arcsec)} & \colhead{($10^{13}~M_{\odot}$)}}
\startdata
\xmmjc\tablenotemark{b} & 7.6$^{+0.8}_{-0.8}$ & 16$^{+1}_{-1}$ & 0.83$^{+0.03}_{-0.03}$ & 
28.6$^{+1.0}_{-0.9}$ & 0.7 & 38.8$^{+2.0}_{-2.7}$ & 1.40$^{+0.14}_{-0.20}$ \\
 & & & & & & & \\
\clA\tablenotemark{a}   & $6.5^{+0.9}_{-0.8}$   &
$10^{+1}_{-1}$ & 
$2.25^{+0.14}_{-0.14}$ & $10.9^{+0.4}_{-0.4}$     & 0.7 &
$39.4^{+2.9}_{-2.9}$  & $1.10^{+0.08}_{-0.08}$ \\ %& $1.2\pm^{0.3}_{0.2}$ & $0.083\pm^{0.013}_{0.011}$\\
 & & & & & & & \\
\rxje\tablenotemark{a}  & $6.6\pm^{1.5}_{1.2}$   & 
$3.6^{+0.4}_{-0.4}  $ & 
$1.14^{+0.12}_{-0.09}$ & $13.2^{+0.9}_{-0.9} $   & $0.7$                  &
$32.1^{+3.6}_{-4.1}$  & $0.66^{+0.09}_{-0.10} $ \\ %& $1.1\pm^{0.4}_{0.4} $ & $0.058\pm^{0.015}_{0.010}$ \\
 & & & & & & & \\
\rxjd\tablenotemark{a}  & $4.5^{+1.5}_{-0.9}$   & 
$1.7^{+0.2}_{-0.2}   $ & 
$0.65^{+0.09}_{-0.08}$ & $17.9^{+3.0}_{-1.7} $   & $0.7$                  &
$26.6^{+4.2}_{-5.2}$ & $0.35^{+0.08}_{-0.10} $ \\ %& $0.53\pm^{0.30}_{0.25} $ & $0.067\pm^{0.023}_{0.014}$ \\
 & & & & & & & \\
\rxjb\tablenotemark{a}  & $6.7^{+2.0}_{-1.5}$   & 
$2.1^{+0.4}_{-0.4}  $ & 
$0.67^{+0.08}_{-0.07}$ & $12.1^{+1.1}_{-1.0} $   & $0.7$                  &
$32.4^{+4.7}_{-5.0}$ & $0.32^{+0.07}_{-0.07} $ \\ %& $1.1\pm^{0.6}_{0.4} $ & $0.029\pm^{0.009}_{0.006}$ \\
 & & & & & & & \\
\xmmjb\tablenotemark{b} & $9.0^{+1.5}_{-1.2}$   & 
$6.9^{+0.4}_{-0.4}$ & 
$1.47^{+0.08}_{-0.08}$ & $12.6^{+0.6}_{-0.5} $   & $0.7$                  &
$36.9^{+3.4}_{-3.6}$  & $0.95^{+0.11}_{-0.12}$ \\ %& $2.0\pm^{0.6}_{0.5} $ & $0.049\pm^{0.010}_{0.008}$ \\
 & & & & & & & \\
\xmmj\tablenotemark{b}  & $7.4\pm^{2.1}_{1.4}$  & 
$2.2\pm{+0.1}  $ & 
$0.58\pm^{0.05}_{0.04}$ & $19.6\pm^{1.2}_{1.4} $   & $0.7$                  &  
$27.5\pm^{3.3}_{3.5}$ & $0.38\pm{0.09}$ \\ %& $0.048\pm^{0.002}_{0.047} $ & $0.069\pm^{0.072}_{0.021}$\\
\enddata
\tablenotetext{a}{0.7-7keV used for spectral analysis}
\tablenotetext{b}{0.5-7keV used for spectral analysis}
\end{deluxetable*}

\subsection{\chandra\ and \xmm\ data analysis}
We analyzed \chandra\ and \xmm\ observations for  each cluster 
that has archival data.

The \chandra\ event files were reprocessed in CIAO~4.1 in order to apply the
latest calibration available (CALDB~4.1). Periods of high background were
excised following the prescription of~\cite{markevitch2003}.
A peripheral region 60-120\arcsec\ from the cluster center was used to determine
the local background; this region allows the background to be determined from
the same chip as the cluster, given the limited angular size
of the sources. %(Fig.~\ref{fig:sx}). 
This choice minimizes the effect of the temporal and spatial variability
 of the Galactic soft X-ray emission \citep[e.g.,][]{snowden1997}.

For the spectral analysis of each cluster, we extracted individual spectra 
and matching response files from each observation separately
(Table~\ref{tab:chandra-data}).

The cluster \xmmjc\ was detected serendipitously in two pointed \xmm\ observations
\citep{lamer2008}, and these observations are the only available X-ray data for this 
cluster. 
We analyze the longest of the two observations, in which the cluster was
 detected in two of the three detectors, MOS1 and PN 
(Table \ref{tab:chandra-data}). For cluster \xmmj\ we analyze both the \chandra\
data and the three longest observations with \xmm.

The \xmm\ data were reduced using the \it{SAS 9.0}\rm~software and the
calibration data available as of July 2009,
and according to the method described
in \citet{nevalainen2005}.
In particular, periods of high background that affected the second
half  of the observation were excluded. We used a local background 
as measured in a peripheral region of each detector, similar
to the method used for the \chandra\ data.

For the purpose of mass calculation and comparison to known scaling relations, we elected
to only use clusters for which archival data are available, and with at least 300 source 
photons.
This selection leaves us with seven clusters: \xmmjc, \clA, \rxjd, \rxjb, \rxje, 
\xmmjb\ and \xmmj. 

\subsection{Image analysis}
\label{subsec:xrayimage}

Event files for each cluster were merged if more than one
observation was available, and images extracted using photons
in the 0.7-7~keV band for ACIS-I observations, and 
in the 0.5-7~keV band for ACIS-S, EPIC-MOS and EPIC-PN observations (see Table~\ref{tab:chandra-data}).
The same table also presents the number of source photons,
after subtracting the expected number of background photons
from the peripheral region.

The gas density is described using an isothermal $\beta$ 
model which, given the limited number of source photons, provides 
a good fit to all clusters with X-ray data:
\begin{equation}
n_{e}(r)=\frac{n_{e,0}}{\left[1+\left(r/r_{c}\right)^{2}\right]^{3\beta/2}}
\end{equation}
Model parameters $n_{e,0}$ and $r_{c}$ are constrained using 
a Markov chain Monte Carlo method described in \citet{bonamente2004}, 
and are presented in Table \ref{tab:xray}.
We fix $\beta=0.7$ throughout. 
Use of the beta model permits a direct comparison with the scaling relations 
presented in B08, obtained using the same isothermal model.

\subsection{Spectral analysis}
\label{sec:spectra}
The spectra for each cluster were extracted from a circular region 
about the centroid of the X-ray emission given in Table~\ref{tab:ubertable}.
A radius of $<$30\arcsec\ was used for all clusters except \xmmjc\ 
and \clA, for which we use $<$60\arcsec; the background spectrum was extracted 
from the surrounding 60-120\arcsec\ region. 
Given the limited S/N of the spectra, the metal abundances were 
fixed at a fiducial value of $A=0.3$\Zsun~for all clusters.
This approximation has a negligible impact on the results of our analysis.

We performed spectral fits to an optically thin model using the APEC
emissivity code \citep{smith2001}; the redshift, Galactic HI column density 
and solar abundance are fixed for each cluster (Table~\ref{tab:chandra-data}), 
leaving just the electron temperature and a normalisation constant.
The resulting electron temperatures are presented in Table~\ref{tab:xray}.

\subsection{Mass measurement}
\label{subsec:xraymass}

The gas model parameters determined from the X-ray images and spectral 
constraints on the gas temperature are used to measure the X-ray gas mass \MgasX.
This is calculated via a spherical integration of $n_{e}$ out to $r_{2500}$ for 
each sample in the Markov chain; this choice 
of radius allows a comparison of $Y$ and \MgasX\ to the scaling relations 
of B08.
The values of $r_{2500}$ and \MgasX\ of each cluster are shown in Table \ref{tab:xray},
with the comparison to $Y$ and previously measured scaling relations in 
Figure~\ref{fig:scaling}.

\section{Results and Discussion}
\label{sec:results}

The \sza\ observations presented here demonstrate the efficacy of using the
SZ effect as a cluster mass discriminator, independent of redshift. 
The SZ effect of the high mass clusters in the ad hoc sample of $z >1$ 
clusters was detected with SZA integration times comparable to those required
 for similar mass clusters at low redshifts.  
Specifically, the SZ effect was detected in the three clusters for which  
\MgasX$\gtrsim10^{13}$\Msun\ and temperature $\geq6.5~\mathrm{keV}$ as 
determined independently from X-ray data (see Table~\ref{tab:xray}). The most distant 
cluster detected by the \sza\ is \xmmjb\ at $z=1.39$, which has a total mass within $r_{500}$ of 
$M_{500}=4.4\pm1.0\times10^{14}$\Msun~\citep{rosati2009}.  
Weak lensing observations by~\cite{jee2009} indicate a total mass within 1 Mpc of 
$8.3\pm1.7\times10^{14}$\Msun.
Similar masses are found for \clA\ ($M_{500}=5.2^{+1.0}_{-0.8}\times10^{14}$\Msun,~\citealt{maughan2006}) 
and \xmmjc\ ($M_{500}=5.6\pm1.0\times10^{14}$\Msun,~\citealt{lamer2008}).

The lack of SZ detections for the other clusters strongly indicates they
 are lower mass systems; in particular, those originally discovered in the 
infrared have $M_{gas} < 5\times10^{12}$\Msun (see Table~\ref{tab:ubertable}) and
their null detection prevents investigation of optical-SZ scalings.

\begin{figure}[h]
\resizebox{\columnwidth}{!}{
\includegraphics{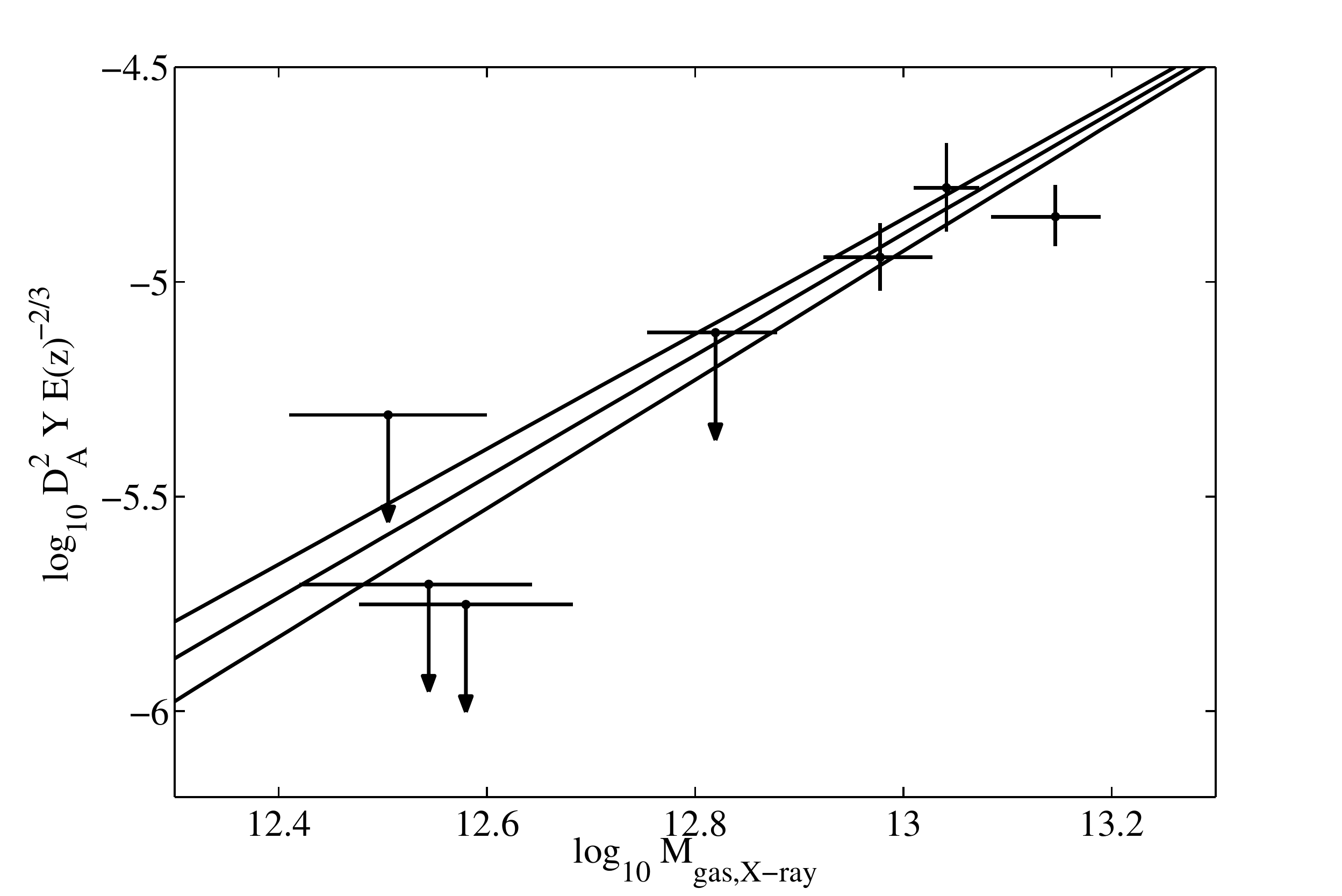}
}
\caption{Comparison of the $Y$ upper limits and detections to \MgasX\
within $r_{2500}$, assuming self-similar evolution. The solid lines are the scaling
 relation measured at $r_{2500}$ by B08, and its $1\sigma$ uncertainties.
In order of X-ray gas mass, the clusters are \rxjb , \rxjd , \xmmj , \rxje , \xmmjb ,
\clA , and \xmmjc.
\label{fig:scaling}}
\end{figure}

As a first step in investigating the SZ--mass scaling relationship at high redshift, we plot 
in Figure~\ref{fig:scaling} the integrated Compton $Y$ values against the X-ray
 gas mass determinations, assuming self-similar evolution. 
The clusters plotted include only those with robust X-ray gas mass constraints 
(see Table~\ref{tab:xray}). 
For comparison with low redshift clusters, the solid lines in Figure~\ref{fig:scaling} 
show the $Y$--\MgasX\ scaling relationship presented in B08 and its $1\sigma$ 
uncertainties.  
The figure illustrates that there is good agreement between the scaling of the high-z
 clusters and that found in the low redshift sample. 
Measurements of more clusters are needed, however, to make a more definitive comparison. 
Ongoing SZ surveys from instruments such as ACT~\citep{fowler2007} and SPT~\citep{carlstrom2009}
 will provide much larger samples of SZ-selected clusters at high redshift \citep[e.g.,][]{vanderlinde2010}.

\section*{Acknowledgments}
The operation of the SZA is
supported by NSF through grant
AST-0604982 and AST-0838187.  Partial support is also provided from grant PHY-0114422 at the University of Chicago, and by NSF grants AST-0507545
and AST-05-07161 to Columbia University.  CARMA operations are supported by the NSF
under a cooperative
agreement, and by the CARMA partner universities.  SM acknowledges support from an NSF
Astronomy and Astrophysics Fellowship; CG, SM, and MS from NSF
Graduate Research Fellowships; DPM from NASA Hubble Fellowship grant HF-51259.01.

\clearpage

\end{document}